\documentclass[a4paper,prd,nofootinbib,preprint,showpacs]{revtex4}

\usepackage[english]{babel}
\usepackage{amsmath,amssymb}
\usepackage[dvips]{graphicx}

\allowdisplaybreaks

\DeclareMathAlphabet{\mathsc}{OT1}{cmr}{m}{sc}

\newcommand{\CL} {C.L. }
\newcommand{\dof}{d.o.f.}

\newcommand{\Sol}  {\mathsc{sol}}

\newcommand{\Dms}{\Delta m^2_\Sol}

\newcommand{\Dcq}{\Delta\chi^2}

\newcommand {\ignore}[1]{}

\def\lsim{\mathrel{\rlap{\lower4pt\hbox{\hskip1pt$\sim$}}
    \raise1pt\hbox{$<$}}}         
\def\gsim{\mathrel{\rlap{\lower4pt\hbox{\hskip1pt$\sim$}}
    \raise1pt\hbox{$>$}}}         

\begin{document}

\raisebox{8mm}[0pt][0pt]{\hspace{12cm}\vbox{hep-ph/yymmdd\\IFIC/02-34}}

  
\title{Confronting Spin Flavor Solutions of the Solar Neutrino
  Problem with current and future solar neutrino data}

\author{J. Barranco $^2$}
\email{jbarranc@fis.cinvestav.mx}
\author{O. G. Miranda$^2$}
\email{Omar.Miranda@fis.cinvestav.mx}

\author{T. I. Rashba$^3$}
\email{rashba@izmiran.rssi.ru}
\author{V. B. Semikoz$^3$}
\email{semikoz@orc.ru, semikoz@ific.uv.es}
\author{J.~W.~F. Valle$^1$}
\email{valle@ific.uv.es}

\address{$^1$Instituto de F\'{\i}sica Corpuscular -- C.S.I.C.,
  Universitat de Val{\`e}ncia \\
  Edificio Institutos, Apt.\ 22085, E--46071 Val{\`e}ncia, Spain}

\address{$^2$Departamento de F\'{\i}sica, Centro de Investigaci{\'o}n y de 
  Estudios Avanzados Apdo. Postal 14-740 07000 Mexico, DF, Mexico}
  
  \address{$^3$ Institute of Terrestrial Magnetism,
    Ionosphere and Radio Wave Propagation of the Russian Academy of Sciences,
    142190, Troitsk, Moscow region, Russia}

\begin{abstract}
  A global analysis of spin-flavour precession (SFP) solutions to the
  solar neutrino problem is given, taking into account the impact of
  the full set of latest solar neutrino data, including the recent SNO
  data and the 1496--day Super-Kamiokande data.  These are
  characterized by three effective parameters: $\Dms \equiv \Delta
  m^2$, the neutrino mixing angle $\theta_\Sol \equiv \theta$ and the
  magnetic field parameter $\mu B_\perp$.
  For the latter we adopt a self-consistent magneto-hydrodynamics
  field profile in the convective zone and identify an optimum
  $B_\perp \sim 80$ KGauss strength for $\mu = 10^{-11}$ Bohr
  magneton.
  We find that no LOW-quasi-vacuum or vacuum solutions are present at
  3~$\sigma$.
  In addition to the standard LMA oscillation solution, there are two
  SFP solutions, in the resonant (RSFP) and non-resonant (NRSFP)
  regimes.  These two SFP solutions have goodness of fit 84 \% (RSFP)
  and 83 \% (NRSFP), slightly better than the LMA oscillation solution
  (78 \%). We discuss the role of solar anti-neutrino searches in the
  fit and present a table of best-fit parameters and
  $\chi^2_{\text{min}}$ values.
  Should KamLAND confirm the LMA solution, the SFP solutions may at
  best be present at a sub-leading level, leading to a constraint on
  $\mu B_\perp$. In the event LMA is not the solution realized in
  nature, then experiments such as Borexino can help distinguishing
  LMA from the NRSFP solution and the simplest RSFP solution with no
  mixing.
    
  In the appendix, we present an updated analysis combining the latest
  data from all solar neutrino experiments with the first results from
  KamLAND. We show that, although the SFP hypothesis still gives an
  excellent description of the solar data, it fails to account for the
  suppressed reactor neutrino flux detected at KamLAND. The inclusion
  of KamLAND excludes the SFP hypothesis at more than 3$\sigma$.
\end{abstract}  

\maketitle

\section{Introduction}
\label{sec:introduction}

The recent neutral current and day-night measurements at the Sudbury
Neutrino Observatory (SNO)~\cite{Ahmad:2002jz,Ahmad:2002ka} as well as
the 1496--day solar neutrino data from
Super-Kamiokande~\cite{Fukuda:2002pe} combined with previous solar
neutrino data~\cite{sun-exp} have shed more light on the long-standing
problem posed by the solar neutrino anomaly.
Evidence from atmospheric neutrino data also indicates that
atmospheric neutrino conversions take place~\cite{Fornengo:2000sr} and
involve mainly two flavors, in order to comply with combined
constraints from reactor neutrinos~\cite{Apollonio:1999ae}.

While the oscillation interpretation of the atmospheric data is rather
robust~\cite{Fornengo:2001pm} present solar data are not yet enough to
pin down the mechanism underlying the neutrino flux suppression.
Although neutrino oscillations provide the most commonly analysed
solution to both solar and atmospheric
anomalies~\cite{Gonzalez-Garcia:2001sq,Maltoni:2002ni} at least two
alternative mechanisms based on non-standard neutrino matter
interactions~\cite{Guzzo:2001mi} and neutrino spin flavor precession
~\cite{Miranda:2001hv,Miranda:2001bi,Akhmedov:2000fj,Guzzo:1999sb,Derkaoui:2001wx}
have been considered.

In this paper we re-consider the status of two-flavor spin flavor
precession (SFP) solutions of the solar neutrino anomaly.
The latter require the existence of non-zero transition magnetic
moments of neutrinos~\cite{Schechter:1981hw} and the interplay of
matter effects~\cite{Akhmedov:1988uk}. 
On general grounds one can argue that, if present, neutrino magnetic
moments should be of this type, as one expects neutrinos to be
Majorana particles~\cite{Schechter:1980gr}.
Such spin flavor precession conversions represent an attractive way of
accounting for present solar neutrino data.
Since they involve only active neutrinos \footnote{Transition magnetic
  moments to sterile neutrinos are rejected by the SNO NC data.} these
are actually the only magnetic-moment-type solutions that survive the
evidence, e.~g.~ from SNO NC data, that solar neutrinos do convert to
\texttt{active} neutrino states, and have just the right features to
reconcile the SNO CC and Super-Kamiokande results.  It has also been
shown how such solutions are robust in the sense that the choice of
the solar magnetic field profile in the convective zone can be made
self-consistently.  Following
refs.~\cite{Miranda:2001hv,Miranda:2001bi} we adopt as profiles the
static magneto-hydrodynamics solutions obtained by Kutvitskii and
Solov'ev~(KS)~\cite{Kutvitsky}.

Here we examine status of all solutions described by the general
two-flavor spin-flavor precession Hamiltonian, which include the
oscillation solutions as a particular case. Using this generalized
picture we show that both SFP solutions give very good descriptions of
the totality of current solar neutrino data, slightly better than the
best oscillation solution, namely LMA~\cite{two}, now robustly
preferred among st the oscillation
solutions~\cite{Maltoni:2002ni,Bahcall:2002hv,Bandyopadhyay:2002xj,Barger:2002iv,deHolanda:2002pp,Strumia:2002rv,Fogli:2002pt}.
For best chosen magnetic field strength there are no vacuum or
LOW-quasi-vacuum solutions at the 3-$\sigma$ level.
However, in addition to the LMA solution, there are two SFP Solutions,
resonant and non-resonant, characterized by goodness of fit 84 \% and
83 \% respectively. We present a table of best-fit values and
$\chi^2_{\text{min}}$ and discuss the role of electron anti--neutrinos
from the sun~\cite{Barbieri:1991ed,Vogel:1999zy,smy} in the fit.
The latter will play an even more important role in the future, should
KamLAND confirm the LMA oscillation solution. In this case the SFP
solution may be realized at best at a sub-leading level, leading to a
constraint on $\mu B_\perp$. This bound is complementary to bounds on
neutrino magnetic moments, such as the one discussed in
ref.~\cite{gri02}. In the event this is not the solution realized in
nature, then future experiments such as Borexino can help
distinguishing between SFP and LMA solutions.

\section{Neutrino Evolution and Conversions}
\label{sec:neutr-conv-prob}

The recent SNO data strongly support that solar neutrinos convert to
\texttt{ active} neutrinos. On the other hand the combined constraints
from reactor neutrino experiments~\cite{Apollonio:1999ae} and
atmospheric neutrino data~\cite{Gonzalez-Garcia:2001sq} imply that
solar neutrino conversions involve mainly two flavors.

Here we consider the evolution Hamiltonian describing a system of two
flavors of active Majorana neutrinos first considered
in~\cite{Schechter:1981hw},
%
\begin{equation}
i\left(
\begin{array}{l}
\dot{\nu}_{eL} \\
\dot{\bar{\nu}}_{eR} \\
\dot{\nu}_{\mu L} \\
\dot{\bar{\nu}}_{\mu R}
\end{array}
\right) = \left(
\begin{array}{cccc}
V_e -c_2\delta & 0 & s_2\delta & \mu B_+(t) \\
0 & - V_e - c_2\delta & - \mu B_-(t) & s_2\delta \\
s_2\delta & - \mu B_+(t) & V_{\mu} + c_2\delta & 0 \\
\mu B_-(t) & s_2\delta & 0 & - V_{\mu} + c_2\delta
\end{array}
\right) \left(
\begin{array}{c}
\nu_{eL} \\
\bar{\nu}_{eR} \\
\nu_{\mu L} \\
\bar{\nu}_{\mu R}
\end{array}
\right)~,  \label{master}
\end{equation}
In eq.~(\ref{master}) $c_2 = \cos 2\theta$, $s_2 = \sin 2\theta$,
$\delta = \Delta m^2/4E$, assumed to be always positive, are the
neutrino oscillation parameters; $\mu$ is the neutrino transition
magnetic moment; $B_{\pm} = B_x \pm iB_y$, are the magnetic field
components orthogonal to the neutrino momentum; $V_e(t) =
G_F\sqrt{2}(N_e(t) - N_n(t)/2)$ and $V_{\mu}(t) =
G_F\sqrt{2}(-N_n(t)/2)$ are the neutrino vector potentials for
$\nu_{eL}$ and $\nu_{\mu L}$ in the Sun, given by $N_e(t)$ and
$N_n(t)$, the number densities of the electrons and neutrons,
respectively.
This generalized form takes into account that, in addition to mixing,
massive Majorana neutrinos may be endowed with a non-zero transition
magnetic moment. 
In the limit where $\mu B \to 0$ this system reduces to the widely
discussed case of two-flavor oscillations.
On the other hand when the mixing vanishes, $\sin 2\theta \to 0$, one
recovers the pure magnetic solutions considered in
\cite{Miranda:2001bi,Akhmedov:2000fj,Guzzo:1999sb,Derkaoui:2001wx}.
As we will see, it will be important to take into account the effects
of neutrino mixing in the characterization of the SFP solutions.

In our calculations of SFP neutrino survival probabilities we use the
electron and neutron number densities from the BP00 model \cite{BP00}
with the magnetic field profile obtained in ref.~
\cite{Miranda:2001bi} for k=6 and $R_0 = 0.6 R_\odot$. Finally, in
order to obtain Earth matter effects we integrate numerically the
evolution equation in the Earth matter using the Earth density profile
given in the Preliminary Reference Earth Model (PREM) \cite{PREM}.


The combined amplitude for a solar $\nu_e$ to be detected as
$\nu_\alpha$ ($\alpha$ being e, $\mu$, $\bar{e}$, $\bar{\mu}$) with
energy $E$ at a detector in the Earth can be written as:
\begin{equation} 
A^{\text{S-V-E}}_{\nu_e\to\nu_\alpha} 
= 
\langle \nu_\alpha | U^{Earth}U^{Vacuum}U^{Sun}| \nu_e \rangle 
= 
\sum_{i=1,2,\bar{1},\bar{2}}A^S_{e\,i}\,A^E_{i\,\alpha}\,\exp[-im_i^2 
(L-R_\odot)/2E]~
\,. 
\label{amplitud} 
\end{equation} 

Here $A^S_{e\,i}$ is the amplitude of the transition $\nu_e \to \nu_i$
($\nu_i$ is the $i$-mass eigenstate) from the production point to the
Sun surface, $A^E_{i\,\alpha}$ is the amplitude of the transition
$\nu_i \to \nu_\alpha$ from the Earth surface to the detector, and the
propagation in vacuum from the Sun to the surface of the Earth is
given by the exponential, where $L$ is the distance between the center
of the Sun and the surface of the Earth, and $R_\odot$ is the radius
of the Sun.  While the presence of magnetic field couples the four
states in the evolution, its absence in vacuum and in the Earth
produces the decoupling of the four states into two doublets : (
$\nu_e$, $\nu_{\mu}$ ) and ($\nu_{\bar{e}}$, $\nu_{\bar{\mu}}$). The
corresponding probabilities $P_{e \alpha}$ can be found by numerically
solving the evolution equation~(\ref{master}).  Note that in the limit
$B_\perp \to 0$ we recover the oscillation case.  

\begin{figure}
\includegraphics[height=6cm,width=15cm,angle=0]{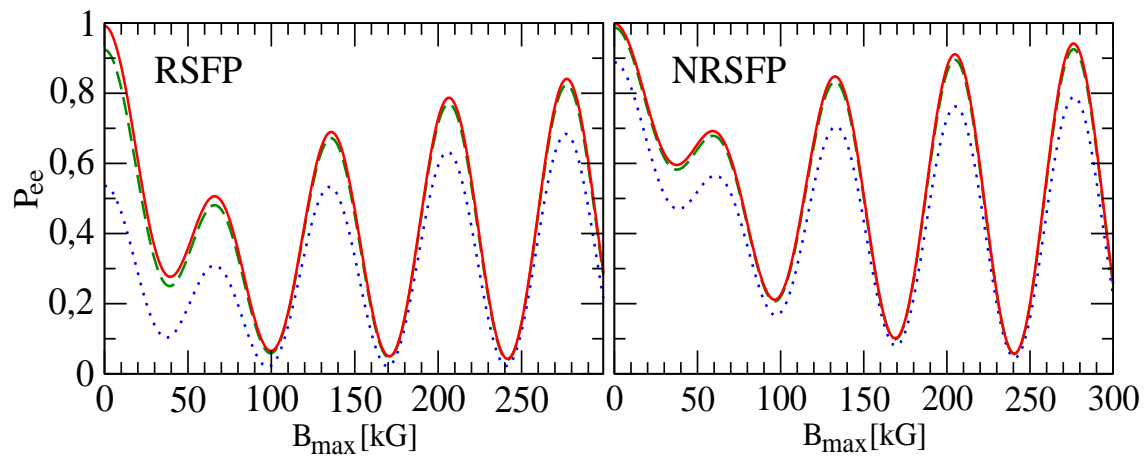}
\caption{Neutrino survival probabilities versus magnetic field strength
  ($B_{\mbox{max}}$).  The neutrino oscillation parameters have been
  fixed as $E/\Delta m^2=1.25\cdot10^8 \mbox{MeV}/\mbox{eV}^2$,
  $\tan^2\theta=0.001$ (solid red line), $0.01$ (green dashed line),
  $0.1$ (dotted blue line) in the left panel (``light'' side).  The
  corresponding numbers for the ``dark''-side (right panel) are
  $E/\Delta m^2=2.5\cdot10^8 \mbox{MeV}/\mbox{eV}^2$,
  $\tan^2\theta=10$ (dotted blue line), $100$ (green dashed line),
  $1000$ (solid red line).  Here we fix $\mu = 10^{-11}$ Bohr
  magneton.}
    \label{fig:probfd}
 \end{figure}

With the above we proceed to analyse the behavior of our neutrino
survival probabilities with respect to variations in $\mu B_\perp$.
For the simple case of constant matter density and field strength and
neglecting Earth regeneration effects, this was given explicitly in
eq.~(8) of ref.~\cite{Miranda:2001hv}. The basic feature to note in
this case is that the survival probability exhibits a periodic
behavior. In Fig.~\ref{fig:probfd} we show that such behavior also
holds in the case of realistic matter density and magnetic field
profiles obtained from magneto-hydrodynamics~\cite{Kutvitsky}.
%
%
 One can see that, both for ``light'' side (RSFP-like) and
 ``dark''-side (NRSFP-like) solutions the neutrino survival
 probabilities exhibit an approximately periodic behavior with
 respect to $\mu B_\perp$.  Note that we have fixed a transition
 magnetic moment of $10^{-11}$ Bohr magneton, consistent with existing
 present experiments.  Examining Fig.~\ref{fig:probfd} one sees that
 the smallest magnetic field magnitude which leads to a boron neutrino
 survival at the required level lies close to 80 KGauss.
 For simplicity, in what follows we will adopt such optimum strength
 $B_\perp \sim 80$ KGauss, because, being the smallest, it is probably
 the one preferred by astrophysics. It is straightforward to repeat our
 analysis for higher field strengths. This will lead to
 ``recurrences'' in solution-space, i.e. to the existence of
 additional branches of the RSFP-like solutions, such as can be seen,
 for example, from Figs. 4 and 6 in ref.~\cite{Miranda:2001bi}.
 Similarly, there will be additional branches of the ``basic''
 NRSFP-like solution found in ref.~\cite{Miranda:2001hv}.  For
 simplicity we focus, in what follows, on the analysis of the
 ``first'' RSFP and NRSFP-like solutions, and their comparison with
 today's favorite oscillation solution, LMA.

\begin{figure}
\includegraphics[height=6cm,width=15cm,angle=0]{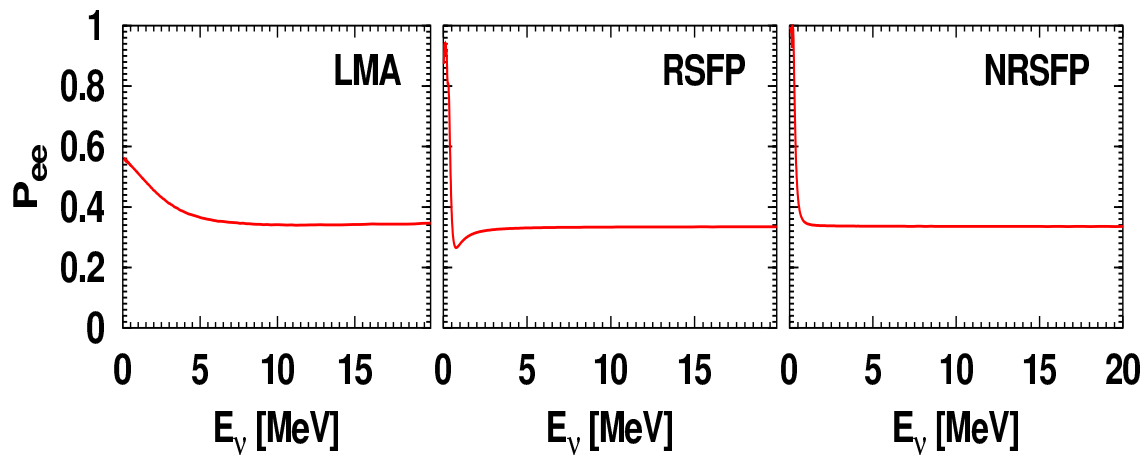}
\caption{
  Optimum two-neutrino survival probabilities for the LMA (left panel)
  and for the two SFP solutions: RSFP (middle panel) and NRSFP (right
  panel).}
    \label{fig:prob}
\end{figure}

In Fig.~\ref{fig:prob} we show a schematic view of the spin flavour
precession survival probabilities for both resonant and non-resonant
cases, RSFP (middle panel) and NRSFP (right panel). The survival
probabilities for the LMA case is also shown, for comparison, in the
left panel.  
For the LMA case the neutrinos are converted into muon neutrinos,
while for the SFP scenario, the neutrinos are mainly converted into
muon anti-neutrinos.
  Note that the above survival probabilities have been given for the
  best-fit parameter values determined in our fit (see table 1,
  below).

\section{Fit procedure}
\label{sec:fit-procedure}

Global analyzes of solar neutrino data have become quite standard, for
a recent reference see~\cite{Maltoni:2002ni}.  Here we briefly
describe the main features of our analysis.

In order to determine the expected event numbers for the various solar
neutrino experiments we calculate the $\nu_e$ survival probability for
each point in parameter space.
We adopt Standard Solar Model neutrino fluxes~\cite{BP00}, treating
however the $^8$B neutrino flux normalization as a free parameter
$f_B$, which is constrained by the SNO NC measurement.
In order to determine the expected signal in each detector, these
fluxes are convoluted with the survival probability at the detector,
the neutrino cross-sections and the detector response functions for
Super-Kamiokande~\cite{Fukuda:2002pe} and SNO~\cite{respsno}.  We use
the efficiencies employed previously, e.~g. in
ref.~\cite{Maltoni:2002ni}. For the SNO case the charged current and
NC cross-sections of neutrinos on deuterium were taken from
ref.~\cite{crossno} and the response functions in ~\cite{respsno}.


We also include theoretical and experimental errors and their
cross--correlations, following the standard covariance approach. In
particular, the errors associated to the energy-scale and the
energy-resolution uncertainties of the Super--Kamiokande and SNO
experiments are included.


Here we use all current solar neutrino data \cite{sun-exp}: the solar
neutrino rates of the chlorine experiment ($2.56 \pm 0.16 \pm
0.16$~SNU), the most recent gallium results SAGE~($70.8
~^{+5.3}_{-5.2} ~^{+3.7}_{-3.2}$~SNU) and GALLEX/GNO ($70.8 \pm 4.5
\pm 3.8$~SNU), as well as the 1496-days Super-Kamiokande data
sample~\cite{Fukuda:2002pe} in the form of 44 bins (8 energy bins, 6
of which are further divided into 7 zenith angle bins).
In addition to this, we include the latest results from SNO presented
in Refs.~\cite{Ahmad:2002jz,Ahmad:2002ka}, in the form of 34 data bins
(17 energy bins for each day and night period). 

Therefore we have in total $3+44+34=81$ observables in our statistical
analysis, which we fit in terms of the parameters $\Dms$,
$\theta_\Sol$. The third parameter $\mu B_\perp$ characterizing the
maximum magnitude of the magnetic field in the convective zone is
fixed at its optimum value $B_\perp = 84$ KGauss. As mentioned, we
employ the self-consistent magneto-hydrodynamics magnetic field
profile obtained in ref.~ \cite{Miranda:2001bi} for k=6 and $R_0 = 0.6
R_\odot$. 


We have compared the data described above with the expected event
numbers, taking into account the relevant detector characteristics and
response functions.  Using a suitable definition of $\chi^2_\Sol$ (the
same as in ref.~\cite{Maltoni:2002ni}, except that we leave the boron
flux free and remove the corresponding theoretical flux errors from
the covariance matrix) we have performed a global fit of present solar
neutrino data.
The allowed regions for a given C.L. are defined as the set of points
satisfying the condition
\begin{equation}
    \chi^2_{\rm SOL}(\Delta m^2,\theta)
    -\chi^2_{\rm SOL,min}\leq \Delta\chi^2 \mbox{(C.L., 2~d.o.f.)} , 
\label{eqn:chirsf}
\end{equation}
where $\Dcq (\text{\CL,\dof})$= 4.61, 5.99, 9.21, 11.83 for 90, 95, 99
\CL and 3$\sigma$, respectively.

\begin{figure}
\includegraphics[height=8cm,angle=0]{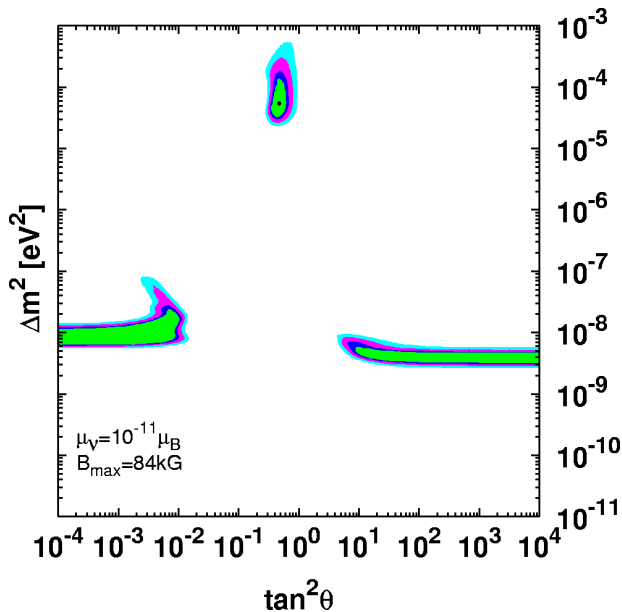}
\caption{
  Allowed regions of $\tan^2\theta_\Sol$ and $\Dms$ for the two-flavour
  spin flavor precession solutions RSFP and NRSFP as well as the LMA
  oscillation solution.}
\label{fig:chi.solsfp}
\end{figure}

We present in Fig.~\ref{fig:chi.solsfp} the allowed regions of
$\tan^2\theta_\Sol$ and $\Dms$ for the two-flavour
spin-flavor-precession.
The confidence levels are $90\%$, $95\%$, $99\%$ and $3\sigma$ for 2
\dof.  As noted previously, we re-confirm the appearance of two new
solutions, totally due to the effect of the magnetic field.  The first
is the RSFP solution \cite{Miranda:2001bi}, which extends up to
$\tan^2\theta$ values around $10^{-2}$ or so. In addition, one finds a
non-resonant (NRSFP) solution \cite{Miranda:2001hv} in the
``dark-side'' of parameter space, for large $\tan^2\theta$ values.
The fits corresponding to these two spin flavor precession solutions
are slightly better than that for the LMA solution, which is recovered
without any essential change due to the effect of the magnetic moment.
Note, however, that the contours are defined with respect to the
global minimum of $\chi^2$, and that this is located at the RSFP
solution. As a result the LMA region in our case is slightly smaller
than the one corresponding to the pure oscillation case (no magnetic
field)~\cite{Maltoni:2002ni}~\footnote{In Fig.~\ref{fig:chi.solsfp} we
  have adjusted the value of $\mu B_\perp$ to its best value (for $\mu
  = 10^{-11} \mu_B$ this corresponds to a maximum magnetic field
  $B_\perp = 84$ KGauss).}.


A characteristic feature of spin flavor precession solutions is that
they produce anti-electron neutrinos, in contrast to the oscillation
case.
One of the new features of our present results is that, in contrast to
what was found previously~\cite{Miranda:2001hv,Miranda:2001bi}, the
solar data alone are now sufficient to rule out all oscillation
solutions other than LMA, without need to include as part of our
$\chi^2_\Sol$ the term corresponding to the data of the LSD
experiment~\cite{Aglietta:1996zu} or the electron anti-neutrino
Super-K flux limits~\cite{smy}.
While the inclusion of these terms would reduce the two SFP branches,
in this paper we will show how the solar neutrino data sample
including the solar anti-neutrino rates at SNO leads to the same
effect.
Before we do that, let us first generalize the neutral and charged
flux discussion given by SNO~\cite{Ahmad:2002jz} to the case where
there is also a third flux, namely that of solar anti-neutrinos,
expected in the SFP scenario.

\begin{figure}
\includegraphics[height=6cm,width=15cm,angle=0]{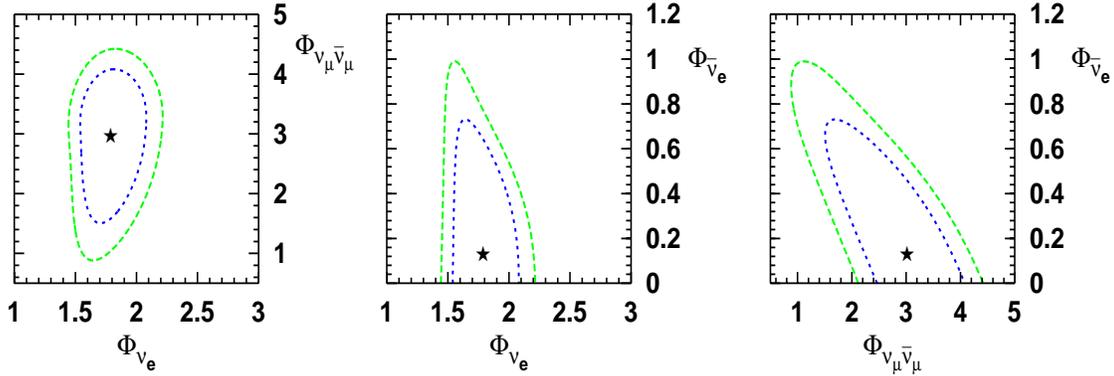}
\caption{
  Solar neutrinos fluxes $\Phi_{\nu_e}$,
  $\Phi_{\nu_\mu\bar\nu_\mu}$=$\Phi_{\nu_\mu}$+$\Phi_{\bar\nu_\mu}$
  and $\Phi_{\bar\nu_e}$ (in units of $10^6 \: \mbox{cm}^{-2}
  \mbox{s}^{-1}$) derived from the number of events per bin at SNO for
  undistorted $^8$B spectrum.  }
\label{fig:NC-CC-anue-fluxes}
\end{figure}

In Fig.~\ref{fig:NC-CC-anue-fluxes} we display the solar neutrinos
fluxes including the electron-antineutrino flux, namely
$\Phi_{\nu_e}$, $\Phi_{\nu_\mu\bar\nu_\mu}$ =
$\Phi_{\nu_\mu}$+$\Phi_{\bar\nu_\mu}$ and $\Phi_{\bar\nu_e}$ as
derived from the observed SNO event number, assuming undistorted $^8$B
spectrum.  Note that fluxes are in units of $10^6 \:\mbox{cm}^{-2}
\mbox{s}^{-1}$ and that here we assume the SSM boron flux prediction.

In the left panel we give the muon neutrino flux versus the electron
neutrino flux.  The middle and right panels we give the electron
anti-neutrino flux versus electron neutrino flux (middle) and versus
the muon-type neutrino flux (right).  These contours correspond to 68
\% \CL and $90\%$ \CL for 2 \dof.  One sees that the electron
anti-neutrino flux, calculated with the cross sections in
\cite{crossno}, is constrained in model-independent way to be less
than $1\cdot 10^6\mbox{cm}^{-2} \mbox{s}^{-1}$ at 90 \% \CL The best
fitted fluxes are: $\Phi_{\nu_e}=1.79 \cdot 10^6 \: \mbox{cm}^{-2}
\mbox{s}^{-1}$, $\Phi_{\nu_\mu\bar\nu_\mu}= 3.02 \cdot 10^6 \:
\mbox{cm}^{-2} \mbox{s}^{-1}$, $\Phi_{\bar\nu_e}= 0.14 \cdot 10^6 \:
\mbox{cm}^{-2} \mbox{s}^{-1}$, with $\chi^2_{\mbox{min}}=16.7$ and 15
\dof. Note that the best fit flux $\Phi_{\bar\nu_e}$ shown by stars in
Fig.~\ref{fig:NC-CC-anue-fluxes} is compatible with zero at $97\%$ \CL

\begin{figure}
\includegraphics[height=8cm,angle=0]{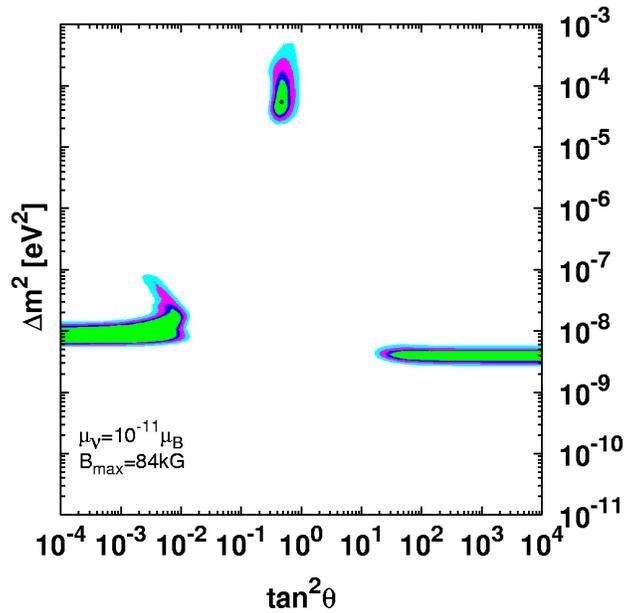}
\caption{
  Same plot as in Fig. \ref{fig:chi.solsfp}, but including electron
  antineutrino CC interaction in SNO ($\bar{\nu_e} + D \to n + n +
  e^+$) with the approximation discussed in text. }
\label{fig:chi:solsfpanue}
\end{figure}

Encouraged by the consistency of our generalization of the SNO
procedure for the SFP case, we now move to the inclusion of the
anti-neutrino data in the analysis.  The reaction $\bar{\nu_e} + D \to
n + n + e^+$ would lead, in addition to the positron Cerenkov light,
to the capture of 2 neutrons by deuteron(s) and the production of two
mono-chromatic gammas with energy $E_\gamma \approx$ 6.25 MeV.
Clearly a complete analysis of the associated signal arising from this
can not be performed at the moment, since we lack the appropriate
response function~\footnote{We thank Art McDonald for useful
  discussions}. As an approximation we can, however, assume that the
positron resolution function is the same as for electrons, and also
that the two neutrons can be captured by Deuterum in the same way as
the neutrons in the NC channel (this last contribution is the most
important one).  Under this approximation we obtain in
Fig.~\ref{fig:chi:solsfpanue} an estimate of the allowed regions.
Particularly noticeable is the fact that in this case the dark side
region becomes smaller. This cut is required in order to avoid an
unacceptably high solar anti-neutrino flux (above 20 \% or so of the
boron-8 neutrino flux) for values of $\tan^2\theta \lsim 20$.  This
exercise highlights the relevance of solar anti-neutrino searches at
SNO in restricting the parameter space of SFP solutions. Such
possibility using Super-K is certainly less favorable.


Before we conclude this section let us present in table~1 the
goodness-of-fit (GOF) corresponding to each of our three solutions.
The analysis has been done for 81 (Chlorine + GNO + SAGE + Zenith SK
spectrum + SNO D/N Spectrum) - 3 parameters: $\tan^2\theta_\Sol$,
$\Dms$, and the boron flux $f_B$, corresponding to 78 d.o.f.  This
table shows the best-fit parameter values, $\chi^2_{\mbox{min}}$ and
GOF for the three solutions. The top panel refers to the usual
oscillation case, the middle panel corresponds to the fit without
including the SNO anti-neutrino CC interaction and the bottom panel is
for the case where this interaction is included. 
\begin{table}
\begin{center}
\begin{tabular}{lllcc}
\hline
\noalign{\smallskip}
Region & $\tan^2\theta$  &  $\Delta m^2$  & $\chi^2_{\mbox{min}}$  & g.o.f.\\
\noalign{\smallskip}
\hline
\noalign{\smallskip}
\multicolumn{5}{c}{standard oscillation}\\
\noalign{\smallskip}
\hline
\noalign{\smallskip}
LMA   & $0.47$            & $5.6\cdot10^{-5}$ & 68.0 & 78\%\\ 
\noalign{\smallskip}
\hline
\noalign{\smallskip}
\multicolumn{5}{c}{$\mu_\nu B\neq0$ without SNO CC $\bar\nu_e$}\\
\noalign{\smallskip}
\hline
\noalign{\smallskip}
RSFP  & $4.2\cdot10^{-4}$ & $7.9\cdot10^{-9}$ & 66.1 & 83\%\\
NRSFP & $119$             & $4.0\cdot10^{-9}$ & 66.4 & 82\%\\
LMA   & $0.47$            & $5.6\cdot10^{-5}$ & 68.0 & 78\%\\
\noalign{\smallskip}
\hline
\noalign{\smallskip}
\multicolumn{5}{c}{$\mu_\nu B\neq0$ with SNO CC $\bar\nu_e$}\\
\noalign{\smallskip}
\hline
\noalign{\smallskip}
RSFP  & $5.3\cdot10^{-4}$ & $7.9\cdot10^{-9}$ & 65.8 & 84\%\\ 
NRSFP & $3\cdot10^{3}$    & $4.0\cdot10^{-9}$ & 66.4 & 82\%\\
LMA   & $0.47$            & $5.6\cdot10^{-5}$ & 68.1 & 78\%\\ 
\noalign{\smallskip}
\hline
\end{tabular}
\end{center}
\caption{Best fit values of $\Delta m^2$ and $\tan^2\theta$ with the
corresponding $\chi^2_{\mbox{min}}$ and GOF for the standard oscillation
case (top panel) and for non-zero magnetic field (middle and bottom panels).  
The squared mass differences are given in eV$^2$.}
\end{table}

\begin{figure}
\includegraphics[height=6cm,width=15cm,angle=0]{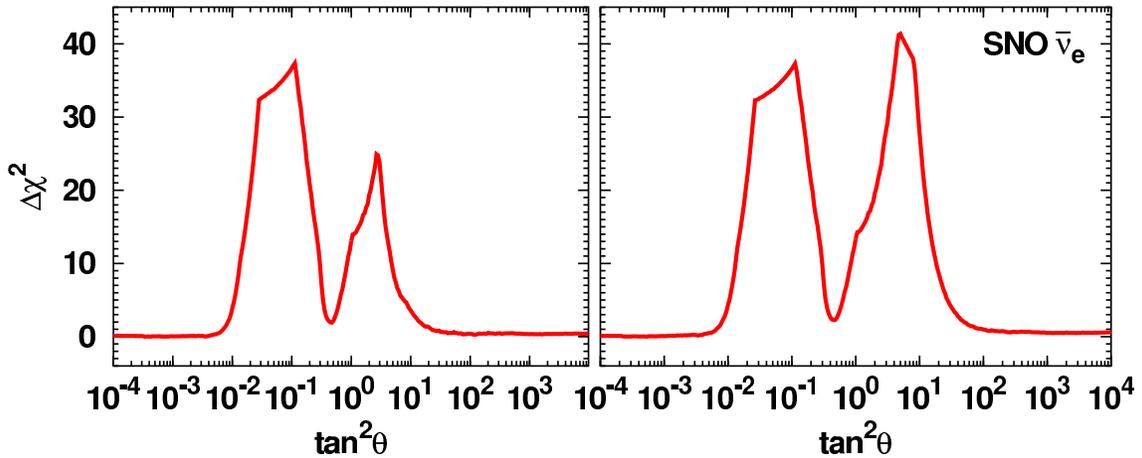}
\caption{
  $\Delta\chi^2_\Sol$ as a function of $\tan^2\theta$ with respect to
  the favored spin-flavor-precession solution}
\label{fig:chi:solsfp}
\end{figure}

To close this session we now illustrate more concisely the above
results by displaying in Fig.~\ref{fig:chi:solsfp} the profiles of
$\Delta\chi^2_\Sol$ as a function of $\tan^2\theta_\Sol$. This is
obtained by minimizing with respect to the undisplayed oscillation
parameters, for the fixed $\mu B_\perp$ value indicated above.
%
Note that the $\Dcq_\Sol$ is calculated with respect to the favored
spin-flavor-precession solution.  The first thing to notice are the
two \texttt{plateaus} corresponding to the RSFP and NRSFP solutions,
slightly lower than the LMA $\chi^2_{\mbox{min}}$.  In contrast to the
left panel, the right panel includes the electron anti-neutrino CC
interaction in SNO.  One can see how the NRSFP \texttt{plateau} has
now become narrower (see also Fig.~\ref{fig:chi:solsfpanue}). Moreover
one can appreciate two very small kinks corresponding to the
``would-be'' LOW solutions. Their status worsens and their position
shifts slightly towards the ``dark side''.

\section{Future Experiments}
\label{sec:future-experiments}

We have seen how the SFP scenario leads to three very good
descriptions of current solar neutrino data corresponding to the RSFP,
NSRFP and the LMA solutions.  The issue comes as to how to distinguish
between these solutions, which are presently statistically equivalent.
Note that this is not such an easy task since, as seen in
Fig.~\ref{fig:prob}, the expected spectral energy distribution for our
spin flavour precession solutions is hardly distinguishable from
the one expected in the pure LMA oscillation solution.  Moreover, in
contrast to the oscillation solution, where a day-night effect is
predicted, the SFP spectra show no day-night asymmetry.

There is at the moment great expectation as to the first physics
results of the upcoming KamLAND experiment, expected
shortly~\cite{kamland}.  Here we analyse the implications of two
possible outcomes of the KamLAND experiment for the status of SFP
solutions.

\subsection{If KamLAND confirms the LMA oscillation solution}
\label{sec:KamLAND}

\begin{figure}
\includegraphics[height=6cm,width=15cm,angle=0]{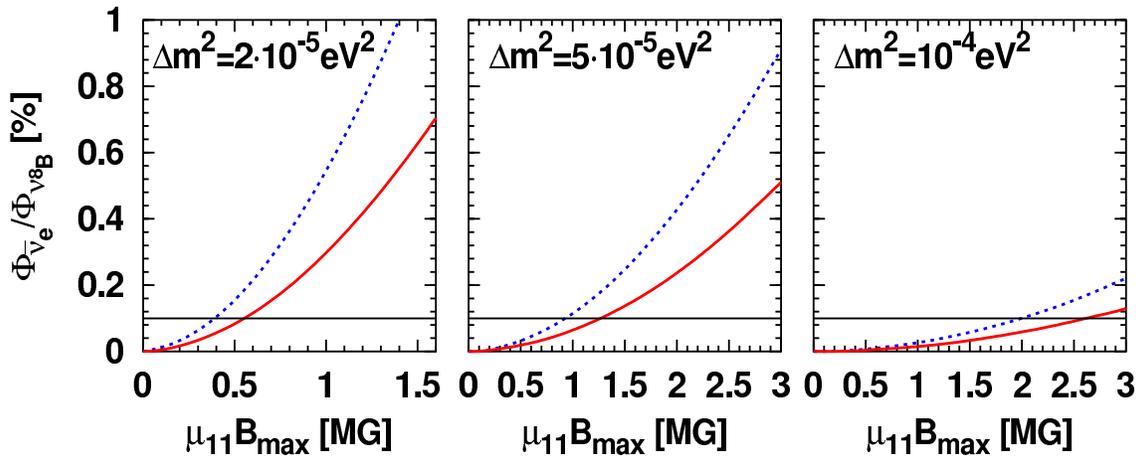}
\caption{KamLAND sensitivity on the Majorana neutrino transition magnetic
  moment in case the LMA solution is confirmed. See text.}
\label{fig:antikam.eps}
\end{figure}

In this case all alternative solutions, such as the present
spin-flavor-precession solutions, may be present only at a sub-leading
level and, on the basis of how good is the KamLAND determination of
the LMA oscillation parameters, one will correspondingly constrain any
exotic alternative.  Our resulting constraints on the SFP solutions
are illustrated in Fig.~\ref{fig:antikam.eps}.

In Fig.~\ref{fig:antikam.eps} we have displayed the electron
anti-neutrino flux predicted at KamLAND ($E>8.3$~MeV) for three
different $\Dms$ values (indicated in the figure) and for
$\tan^2\theta_\Sol$ values varying in the range from 0.3-0.8, as a
function of $\mu_{11} B_{\text max}$, $\mu_{11}$ being the magnetic
moment in units of $10^{-11}$ Bohr magneton and $B_{\text max}$ being
the maximum magnetic field in the convective zone. In order to obtain
such a simple correlation we note the importance of using our
self-consistent magneto-hydrodynamics magnetic field profile, as in
ref.~ \cite{Miranda:2001bi}. The extremes of the neutrino mixing range
correspond to the solid and dashed lines indicated in the figure,
while the horizontal line corresponds to a KamLAND sensitivity on the
anti-neutrino flux of 0.1~\%, expected with three years
running~\cite{kamland}. Clearly the limits on the transition magnetic
moments are sensitive also to the ultimate central $\Dms$ value
indicated by KamLAND (a 10 \% error is expected), being more stringent
for lower $\Dms$ values, as seen from the left panel.

\subsection{If KamLAND does not confirm the LMA solution}
\label{sec:kamland-does-not}

Imagine now the extreme and unlikely case that KamLAND does not
provide useful information on the neutrino parameters. In this case
one can compare the predictions of these three solutions for the
upcoming Borexino experiment, as suggested in
ref.~\cite{Akhmedov:2002ti}.

\begin{figure}
\includegraphics[height=7cm,angle=0]{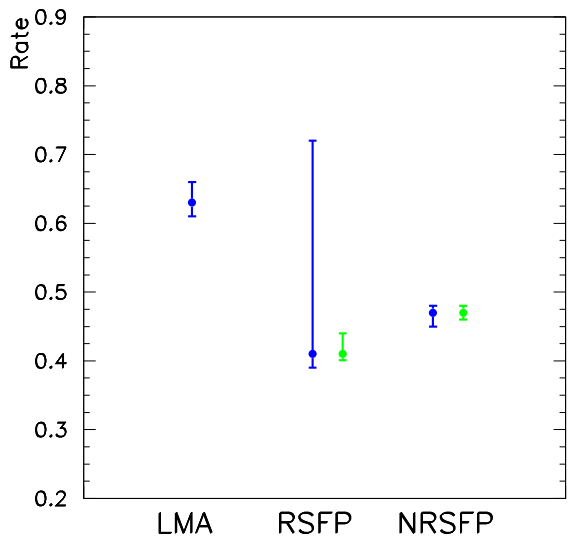}
\caption{Predicted $R_{\rm Borexino}$ values for LMA and
  spin flavor precession solutions.}
\label{fig:boro3.eps}
\end{figure}

\begin{figure}
\includegraphics[height=7cm,angle=0]{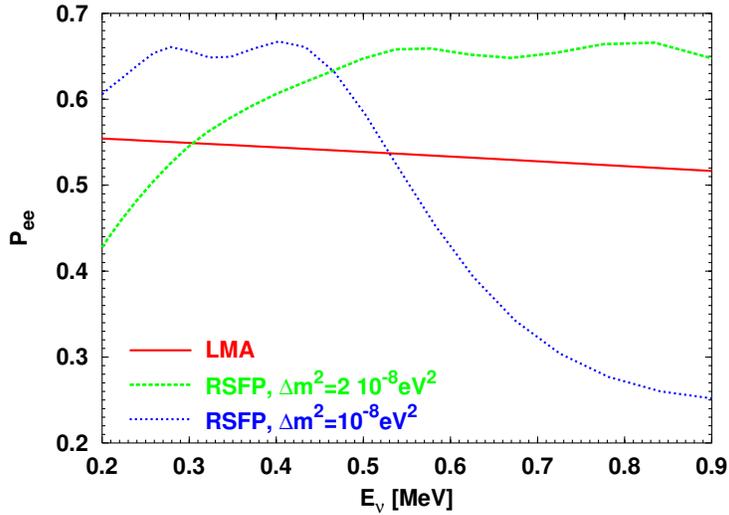}
\caption{Neutrino survival probability for RSFP-like solution 
  with $\Dms \sim 2 \times 10^{-8}$ eV$^2$ and $\tan^2\theta_\Sol
  \sim 10^{-2}$, in the range between 0.2 and 0.9 MeV. }
\label{fig:closeup}
\end{figure}

A simple way to display this is presented in Fig.~\ref{fig:boro3.eps}.
This figure shows the predicted values for
\begin{equation}
  \label{eq:r7}
R_{\rm Borexino} \equiv \frac{\rm observed \: Borexino \: rate} {\rm SSM \: rate} 
  \end{equation}
  defined as the ratio of observed-over-SSM-expected signal in
  Borexino, assuming best-fit parameters as determined in our fit, and
  90\%\CL error bars for LMA (left line), RSFP (middle lines) and
  NRSFP (right lines) solutions. For the SFP solutions two error bars
  are indicated for $R_{\rm Borexino}$.
%
  The error bars indicated in grey (green in color printers) refers to
  the cases $\theta=0$ for the RSFP case and $\theta=\pi/2$ for the
  NRSFP solution.
  The error bars indicated in dark (blue in color printers) correpond
  to the general SFP case with non-zero mixing.
  Clearly, the zero mixing RSFP and the LMA solutions lead to very
  different predictions, as already noted in
  ref.~\cite{Akhmedov:2002ti}. However, we remark that in general the
  RSFP solution can \texttt{not} be distinguished from LMA insofar as
  the Borexino predition is concerned. Indeed, as seen from
  Fig.~\ref{fig:chi:solsfp}, RSFP is characterized by a very flat
  \texttt{plateau} of nearly constant $\chi^2$, while the Borexino
  prediction depends rather strongly on the poorly determined value of
  the ``best'' neutrino mixing angle for this solution.
  To see this let us take, for example $\Dms \sim 2 \times 10^{-8}$
  eV$^2$ and $\tan^2\theta_\Sol \sim 10^{-2}$ and display the neutrino
  survival probability versus energy, as seen in
  Fig.~\ref{fig:closeup}.
One sees from this figure how in the pp region this solution is
similar to LMA, but leads to a much smaller suppression of the
berilium line. This allows us to understand the inability to predict
with precision the borexino rate expected in this part of the (lowest)
RSFP region. In contrast, the NRSFP solution in the ``dark-side'' is
the one which is more clearly distinguishable from the LMA oscillation
solution.  Moreover, we have verified that, for the case of the NRSFP
solution the Borexino predition is rather insensitive to whether the
neutrino mixing is left free or not.

\begin{figure}[t]
\includegraphics[height=6cm,width=15cm,angle=0]{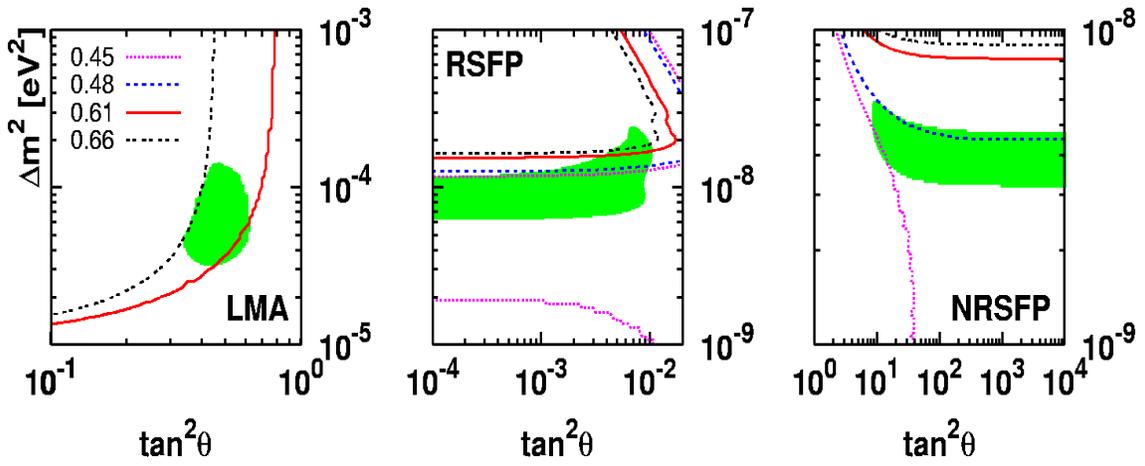}
\caption{Contour lines of predicted $R_{\rm Borexino}$ values 
  as a function of the neutrino oscillation parameters. The shaded
  regions correspond to the three solutions discussed here (LMA, left
  panel), RSFP (middel panel) and NRSFP (right panel). See text.}
\label{fig:xi84borex.bmp.eps}
\end{figure}

  A more complete way to present this information is displayed in
  Fig.~\ref{fig:xi84borex.bmp.eps}. Here we have overlapped the
  allowed neutrino parameter ranges determined from our fit with the
  predicted $R_{\rm Borexino}$ values, for each one of these
  solutions, RSFP, NRSFP and LMA at 90\%\CL The results for the LMA
  solution agree well with those obtained, say, in refs.
  \cite{Bahcall:2002hv,deHolanda:2002pp}.
  We can clearly identify from Fig.~\ref{fig:xi84borex.bmp.eps} which
  range in the 90 \% allowed confidence RSFP region leads to the large
  ``spread'' in the borexino prediction discussed previously.

\section{Summary and Discussion}
\label{sec:summary-discussion}

In this paper we have re-considered the solutions of the solar
neutrino problem involving two-flavor oscillations and spin-flavour
precessions.  We have given a global analysis of such SFP solutions to
the solar neutrino problem, taking into account the impact of the full
set of latest solar neutrino data, including the recent SNO data as
well as the 1496--day Super-Kamiokande data.  These are characterized
by three effective parameters: $\Dms \equiv \Delta m^2$, the neutrino
mixing angle $\theta_\Sol \equiv \theta$ and the magnetic field
parameter $\mu B_\perp$. For the latter we have fixed $\mu = 10^{-11}$
Bohr magneton, with a corresponding optimized self-consistent
magneto-hydrodynamics magnetic field profile with $B_\perp \sim 80$
KGauss as the maximum field strength in the convective zone.
Two-flavor oscillations are recovered as a particular case of our
general SFP scenario.
We have found that no LOW-quasi-vacuum or vacuum solutions are present
at the 3~$\sigma$ level.
In addition to the standard LMA oscillation solution, we have
re-confirmed the existence of two SFP solutions, in the resonant
(RSFP) and non-resonant (NRSFP) regimes.  These two SFP solutions have
goodness of fit 84 \% (RSFP) and 83 \% (NRSFP), slightly better than
the LMA oscillation solution (78 \%). We have discussed the role of
solar anti-neutrino searches in the fit and present a table of
best-fit parameters and $\chi^2_{\text{min}}$ values.
Should KamLAND confirm the LMA solution, the SFP solutions may at best
be present at a sub-leading level, leading to a meaningful constraint
on $\mu B_\perp$. If the magnetic field strength is known from solar
physics, we can obtain a bound on the Majorana neutrino transition
magnetic moment, complementary to the bounds discussed in
ref.~\cite{gri02}. In the event LMA is not the solution realized in
nature, then future experiments such as Borexino can be of help in
distinguishing LMA from the NRSFP solution, as well as the restricted
RSFP solution with no mixing.

\appendix

\section{Implication of the KamLAND results}

The original version of this paper indicated that both RSFP and NRSFP
descriptions of the solar neutrino data were slightly better than the
favoured LMA-MSW solution.  Since then the results of the first 145.1
days of reactor neutrino observations at the KamLAND experiment have
been published~\cite{Eguchi:2002dm}.  Complementing our discussion in
Section~\ref{sec:future-experiments} here we analyze the implications
of KamLAND results for the neutrino spin-flavor precession hypothsis.


KamLAND is a reactor neutrino experiment whose detector is located at
the Kamiokande site.  Most of the $\overline{\nu}_e$ flux incident at
KamLAND comes from nuclear plants at distances 80-350 km from the
detector, making the average baseline of about 180 km, long enough to
test the LMA-MSW region. The target for the $\overline{\nu}_e$ flux
consists of a spherical transparent balloon filled with 1000 tons of
non-doped liquid scintillator, and the antineutrinos are detected via
the inverse neutron $\beta$-decay process $\overline{\nu}_e + p \to
e^+ + n$. The KamLAND collaboration has for the first time observed
the disappearance of neutrinos produced in a power reactor during
their flight over such distances. The ratio of the observed events to
the expected number is $0.611\pm 0.085 {\rm (stat)} \pm 0.041 {\rm
  (syst)} $ for $\bar{\nu}_e$ energies $>$ 3.4 MeV.  This gives the
first terrestrial confirmation of the solar neutrino anomaly and also
confirms the oscillation hypothesis with man-produced neutrinos.


The global analysis of solar+KamLAND data for pure neutrino
oscillations (neglecting possible sub-leading effects such as
spin-flavor precession) has been given in~\cite{Maltoni:2002aw}.
Assuming CPT invariance the main result is that all other flavor
oscillation solutions such as vacuum oscillations, SMA-MSW and LOW are
in disagreement with KamLAND data at more than 99.73~\% C. L. We
expect that, under the same CPT invariance assumption, all
non-oscillation solutions will also be rejected, as these would not
account for the reduced reactor neutrino flux detected at KamLAND. In
fact this has already been shown for Non-Standard neutrino matter
Interaction hypothesis (NSI)~\cite{Guzzo:2001mi}.


We have used the Poisson distribution in the analysis of the KamLAND
data, adding this to our previous result for the solar data.  Besides
our previous Boron free analysis, we have also made a Boron fixed
analysis.

%
\begin{figure}
\includegraphics[height=8cm,angle=0]{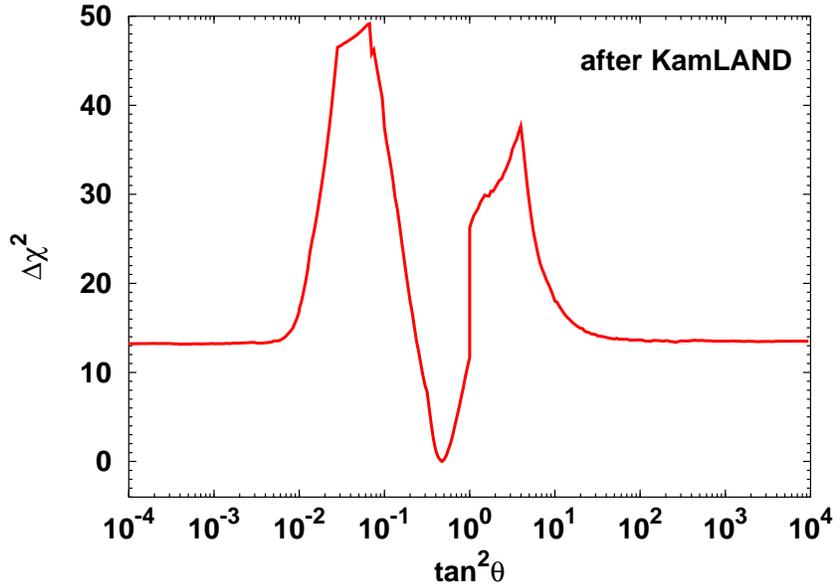}
\caption{$\Delta\chi^2_{\Sol+KamLAND}$ as a function of
$\tan^2\theta$ with respect to the favored LMA solution}
\label{fig:chi:solkamsfp}
\end{figure}
We find that, after adding the KamLAND result, the RSFP and NRSFP
solutions are allowed only at 99.86~\% C. L. and 99.88~\% C. L.,
respectively.  Thus the spin flavor precession solution to the solar
neutrino problem can not be reconciled with the KamLAND data and is
therefore rejected. 
In Fig.~\ref{fig:chi:solkamsfp} the profile of
$\Delta\chi^2_{\Sol+KamLAND}$ as a function of $\tan^2\theta_\Sol$ is
shown. One can see how after KamLAND the RSFP and NRSFP solutions have
now become disfavored in contrast to Fig.~\ref{fig:chi:solsfp} for the
solar data only.
Spin-flavor precessions may however be present at a subleading level.
Future data may be used to place limits on neutrino magnetic moments
and solar magnetic fields.  In particular, for strong solar magnetic
fields and large enough neutrino transition magnetic moment the
LMA-MSW allowed region may be potentially distorted.


\begin{acknowledgments}
  We would like to thank E. Akhmedov, Michele Maltoni and Mariam
  T\'ortola for useful discussions and K. Kubodera, for providing
  anti-neutrino-deuteron cross sections.  This work was supported by
  Spanish grants PB98-0693, by the European Commission RTN network
  HPRN-CT-2000-00148, by the European Science Foundation network grant
  N.~86, by Iberdrola Foundation (VBS) and by INTAS grant YSF
  2001/2-148 and CSIC-RAS agreement (TIR).  VBS and TIR were partially
  supported by the RFBR grants 00-02-16271 and OGM was supported by
  the CONACyT-Mexico grants J32220-E and 35792-E.
\end{acknowledgments}


\end{document}